# Shaping graphene superconductivity with nanometer precision

E. Cortés-del Río [1,2], S. Trivini [3], J.I. Pascual [3,4], V. Cherkez [5,6], P. Mallet [5,6], J-Y. Veuillen [5,6], J.C. Cuevas [2,7,8], I. Brihuega [1,2,8] *

*__Corresponding Author__ *ivan.brihuega@uam.es*

[1] Departamento Física de la Materia Condensada, Universidad Autónoma de Madrid, E-28049 Madrid, Spain.
[2] Condensed Matter Physics Center (IFIMAC), Universidad Autónoma de Madrid, E-28049 Madrid, Spain.
[3] CIC nanoGUNE-BRTA, 20018 Donostia-San Sebastián, Spain
[4] Ikerbasque, Basque Foundation for Science, 48013 Bilbao, Spain.
[5] Université Grenoble Alpes, CNRS, Institut Néel, F-38400 Grenoble, France.
[6] CNRS, Institut Neel, F-38042 Grenoble, France.
[7] Departamento Física Teórica de la Materia Condensada, Universidad Autónoma de Madrid, E-28049 Madrid, Spain.
[8] Instituto Nicolás Cabrera, Universidad Autónoma de Madrid, E-28049 Madrid, Spain

**Abstract:**

Graphene holds great potential for superconductivity due to its pure two-dimensional nature, the ability to tune its carrier density through electrostatic gating, and its unique, relativistic-like electronic properties. At present, we are still far from controlling and understanding graphene superconductivity, mainly because the selective introduction of superconducting properties to graphene is experimentally very challenging. Here, we have developed a method that enables shaping at will graphene superconductivity through a precise control of graphene-superconductor junctions. The method combines the proximity effect with scanning tunnelling microscope (STM) manipulation capabilities. We first grow Pb nano-islands that locally induce superconductivity in graphene. Using a STM, Pb nano-islands can be selectively displaced, over different types of graphene surfaces, with nanometre scale precision, in any direction, over distances of hundreds of nanometres. This opens an exciting playground where a large number of predefined graphene-superconductor hybrid structures can be investigated with atomic scale precision. To illustrate the potential, we perform a series of experiments, rationalized by the quasi-classical theory of superconductivity, going from the fundamental understanding of superconductor-graphene-superconductor heterostructures to the construction of superconductor nanocorrals, further used as "portable" experimental probes of local magnetic moments in graphene.

## 1. Introduction

A simple and effective way to add new properties to a material is to bring it into contact with another material possessing the desired properties, known as the proximity effect. This enables to introduce chosen properties in the neighbouring region of the contact, which can be used to induce magnetic, topological, or superconducting properties among others [1]. A key requirement for an efficient proximitization is the fine control of the junction between the different materials [1-3]. The superconducting proximity effect has been particularly prolific [4-7, 8]. When a normal metal is in contact with a superconductor, Cooper pairs tunnel from the superconductor and penetrate the normal metal forming a 'condensate' of Cooper pairs, which induces superconducting properties on it. The normal metal inherits superconducting correlations (coherence between pairs of electrons of opposite spins) and therefore, can sustain a dissipationless current (or supercurrent), exhibit a diamagnetic response (Meissner effect), etc., i.e., it essentially behaves as an intrinsic superconductor. As a result, the proximity effect is an important tool for understanding the properties of superconducting materials and can help researchers to design and create better superconductors.

By means of the proximity effect, superconductivity was induced in graphene just two years after its discovery [9]. As a result of this, several exciting observations emerged rapidly [10-12]. However, after this promising initial stage, mostly focused on the transport properties of graphene-superconductor hybrids, the experimental progress in the field has been relatively scarce [13-15]. This has likely been due to the difficulties in controlling the preparation of clean samples with well-defined graphene superconductor junctions [14, 15] and the lack of local probe experiments providing more direct information about the superconductivity in graphene [16-18] It was only very recently that graphene superconductivity made a disruptive advance with the appearance of twisted bilayers [19-24].

Here, we present a new method, based on the proximity effect in combination with STM manipulation capabilities [25, 26], that offers an unprecedented control of graphene superconductivity. The manipulation, with nanometre scale precision, of the position of lead islands over the graphene surface with the STM tip, provides a new playground to create S-N-S junctions, where the specific configurations of the superconducting leads become a tuneable parameter.

## 2. Results and Discussion

A key factor in the success of our method lies in executing all the necessary steps (graphene growth, Pb island deposition, and Pb island manipulation) in-situ [27], under the same UHV atmosphere, which ensures a controlled environment. The method works in different types of graphene, ranging from neutral Monolayer (ML), doped ML and bilayers, to graphite. Unless explicitly specified, the results shown here have been obtained on multilayer graphene on SiC(000-1), prepared as indicated in the Methods section. On this surface, graphene behaves as quasi neutral monolayer with a large elastic mean free path [28, 29]. Depositing lead in vacuum and at room temperature resulted in randomly distributed, single-crystal triangular-shaped crystalline Pb islands of nanometre-sizes (typically with 20-300 nm lateral sizes and 2-10 nm heights, see Figure 1a, [17, 30]). At 4K, the base temperature of our scanning tunnelling microscope, the islands maintain the bulk superconductivity, as we probed with differential conductance (*dI/dV*) spectra [17] using Pb SC tips for improving the energy resolution beyond the thermal limit [31, 32]. Scanning tunnelling spectroscopy (STS) results also revealed that graphene regions in the vicinity of the Pb islands develop superconductivity due to proximity effect.

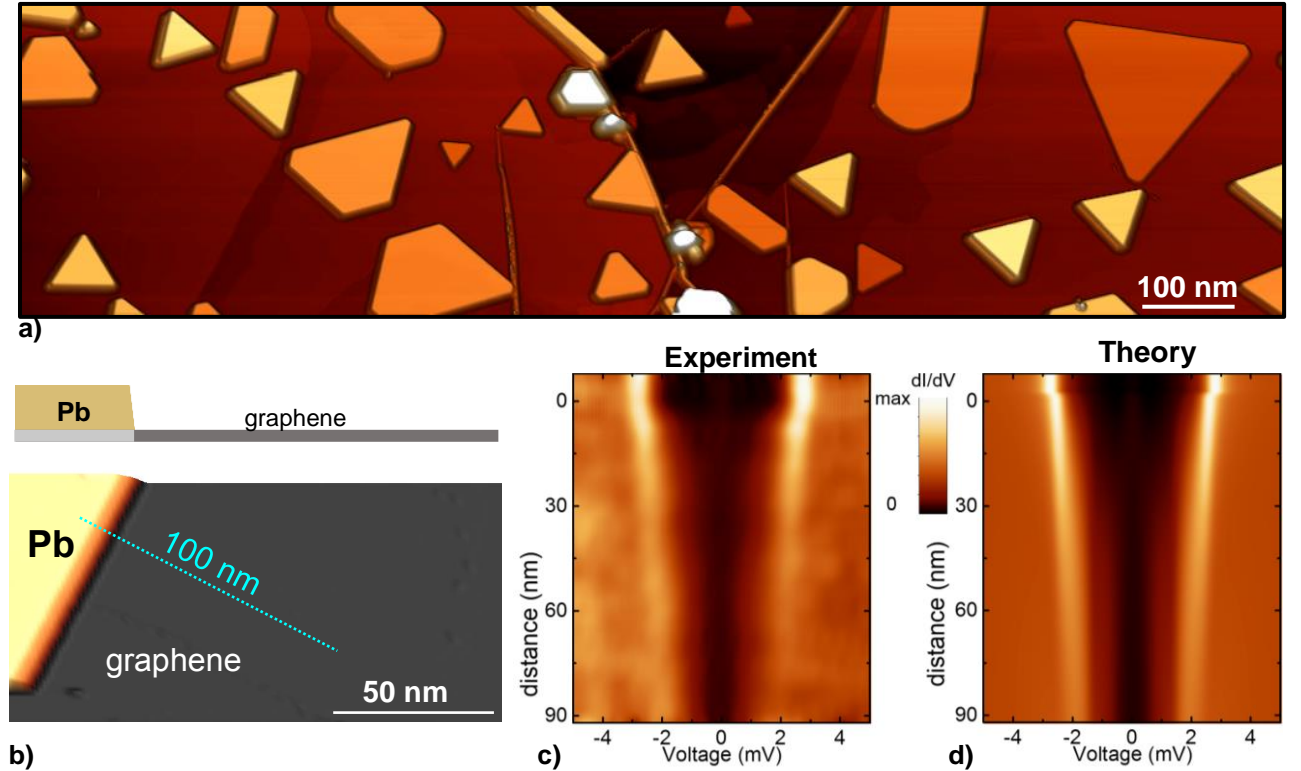

**Figure 1. Inducing SC in graphene by proximity**. Upper panel. Large scale STM image showing the general morphology of the samples after RT Pb deposition. Triangular-shaped Pb islands are formed, at non-controlled positions, on top of the graphene surface grown on SiC(000-1). Temperature = 4.2K; Vbias=1.0 V; It=0.05 nA. a) STM image showing an 8nm height Pb island on top of graphene grown on SiC(000-1). b) Schematic of the system view as a S-N junction, formed by the graphene area below the Pb island, light gray line, as the SC region and the uncovered graphene area, dark gray, as the normal one. c) Experimental conductance map [d$I$/d$V(x,E)$] along the blue dotted line in (a), showing the evolution of SC as a function of *x*, the distance from the Pb island. Negative *x* means *dI/dV* spectra measured on top of the Pb island (*V*bias = 10 mV; *I*set = 0.5 nA). d) Calculated conductance map [d$I$/d$V(x,E)$]. We include the broadening parameters, $\Gamma_G$, which is phenomenological, and $\Gamma_{sf}$, to account for spin-flip scattering mechanisms that can take place due to different types of magnetic scatterers intrinsically present in our graphene samples, such as vacancy, defects, or GB. $T_{sample}$ = 4.3 K; $T_{tip}$ = 3.1 K; $\Delta_G$ = 1.20 meV; $\Gamma_G$ = 0.20 meV; $\Gamma_{sf}$ = 0.2$\Delta_{Pb}$; $\xi$ = 70nm .For the simulations we used the experimentally determined $\Delta_{tip}$ = 1.37 meV and $\Gamma_{tip}$ = 0.08 meV.

To characterize the proximity-induced superconductivity in the graphene around Pb islands, we measured the evolution of the differential conductance as a function of $d$, the distance to the island's edge. Spectral plots of dI/dV vs. $d$ like in Fig. 1c show that superconductivity in graphene persists for tens of nm away from the Pb islands, with a gradual decrease in the intensity and energy separation of the coherence peaks. For the case of Figure 1, the closest spectrum to the island edge is very similar to that of the Pb island itself, which reproduces the local density of states (LDOS) of bulk Pb. On other islands, however, smaller gaps are observed on the graphene close to the island's edges, revealing that the proximitized graphene superconductivity depends strongly on the specific details of the Pb-graphene interface.

To interpret the STS results, we use a quasi-classical theory of superconductivity. The experimental results are consistent with graphene behaving as a diffusive metal, i.e., a regime where the elastic mean free path $l$ is smaller than the superconducting coherence length $\xi$. The proximity effect in this regime is described by the Usadel equations [33, 34], and, in practice, we can disregard any specific detail of graphene's electronic and atomic structure. We analyse our experimental data considering S-N junctions formed by "SC" graphene regions below the Pb islands, whose SC properties are determined by the specific details of each Pb island-graphene interface, and "normal" uncovered graphene regions. From a direct comparison of the experimental dI/dV spectra with the simulations, we can extract both the graphene diffusion coefficient, $D = \frac{1}{2} v_F \cdot l$, where $v_F$ is the graphene Fermi velocity, and the gap of the "SC" graphene region below the island, $\Delta_G$. These quantities define the superconducting coherence length of the graphene layer, $\xi = \sqrt{\frac{\hbar D}{\Delta_G}}$, see Supporting information for more details on the theoretical description of the proximity superconductivity.

The Pb-graphene junction shown in Figure 1 effectively behaves as a one-dimensional problem. Thus, by applying our model to reproduce these measurements, we obtain a diffusion coefficient, $D$ = 89 cm$^2$/s; and an induced gap $\Delta_G$ = 1.20 meV, both resulting in a coherence length $\xi$ = 70 nm. Our experiments show that, for the Pb-graphene junctions in graphene on SiC(000-1), the coherence length is always large, with a value of the order of $\xi \approx$ 70nm. In contrast, the values of $\Delta_G$, obtained from dI/dV spectra close to the islands, varies strongly for different Pb-graphene junctions, ranging from $\Delta_G$ = 1.2 meV, very close to bulk Pb ($\Delta_{Pb}$ = 1.35meV), to $\Delta_G$ = 0.6meV. These $\Delta_G$ values are all significantly larger than the proximity-induced gaps by Pb islands on hydrogen-intercalated single-layer graphene on SiC(0001), where values around $\Delta_G$ = 0.2 meV have been reported [18], indicating that in the present case the Pb-graphene interface has a larger transparency. The typical values employed in our fits for the broadening or Dynes' parameter $\Gamma_G$ are on the order of $0.1\Delta_{Pb}$, which are quite standard for strong-coupling superconductors like Pb[17]. On the other hand, for the spin-flip rate we typically find values $\Gamma_{sf} \approx 0.1 - 0.2\Delta_G$, which implies that the corresponding spin-flip length over which the quasiparticle spin is preserved is of the order of $L_{sf} = \hbar D/\Gamma_{sf} \approx 160 - 230$ nm. This is indeed compatible with our expectations based on the topography of our samples.

We have also investigated different graphene systems, such as monolayer (ML) and bilayer (BL) graphene grown on the SiC(0001), both electron doped, and HOPG surfaces. In all these systems room-temperature evaporation of Pb resulted in the formation of triangular Pb islands, similar to those shown on graphene on SiC(000-1) (refer to Supporting Figures. S1, S2, and refs. [30, 35, 18]). Our results on these surfaces reveal that in all cases bulk-like superconductivity persists in Pb islands, inducing superconducting properties to the surrounding graphene regions. Given the variations in transport properties among different graphene samples of diverse nature and doping [36], the extension of the induced superconductivity varies accordingly. For instance, in the supporting Figure S3, we show the results on HOPG, were the coherence length of the induced superconductivity is found to be $\xi \approx 10$ nm, much shorter than the 70 nm found on SiC(000-1).

Pb islands grow randomly distributed over the graphene sample, exposing distinct regions of the graphene domains to proximitized superconductivity. Here, we present a consistent method to laterally move the Pb islands at desired position of the graphene surface using the STM tip. This enables us to prepare customized superconducting-normal hybrid structures and study their local properties using dI/dV spectra. In Figs 2a-b we show an example of such a manipulation procedure, where we have selectively displaced a triangular Pb island by 60 nm along the graphene surface (indicated by a green arrow). As we sketch in Figure 2c, the manipulation method consists of laterally pushing the Pb islands using the tip apex, which slides over the graphene surface. The methodology is the following. First, we place the STM tip on top of an uncovered graphene region close to the chosen island to move. Second, we open the feedback-loop, set the bias voltage to 0 mV and then move the STM tip parallel to the graphene surface, at a constant speed of 100 nm/s, towards the Pb island. Since, for typical tunnelling feedback parameters, the STM tip apex is closer to the graphene surface than the Pb islands height, during this motion, the STM tip reaches the edge of the Pb island, makes a soft contact with it, and pushes the island to the desired position. The tip is moved until the selected new position for the island is reached. Finally, we retract the tip, letting the Pb island on its final position on graphene and we restore the initial feedback conditions. When pushing the Pb islands with the STM tip a minor trace is left at the at the point where the tip contacts the island, as evident from the small bump at the tip of the green arrow. This minor modification in the island's integrity has no effect on the superconducting properties of the island or on the superconductivity induced in graphene.

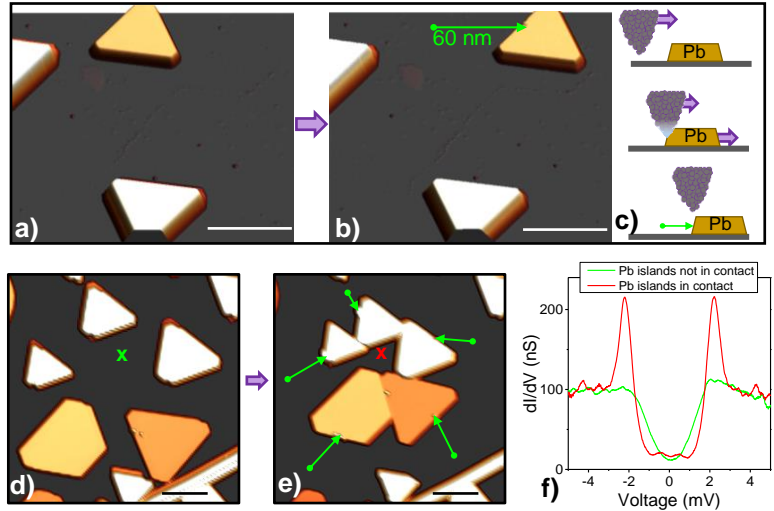

**Figure 2. Pb islands manipulation on graphene**. a-b) Consecutive STM images showing the horizontal manipulation of a SC Pb island. c) Illustration of the manipulation method, by pushing the Pb island with the STM tip. d-e) Collective manipulation of five Pb island to induce a strong SC on the graphene region outlined by a cross, see red and green dI/dV curves in f), measured on this graphene spot, before (green) and after (red) the islands manipulation.

Notice in Figs. 2a-b that the 50 nm wide Pb island moves as a single unit, without rotating with respect to the graphene lattice. As grown Pb islands have their edges parallel with the underlying graphene lattice [30] and maintain this alignment during the lateral movement within the same graphene domain. In contrast, when Pb islands are displaced to another graphene grain, they rotate to align their edges with the underlying graphene domain. We find that, in their new position, Pb islands proximitize the surrounding graphene as they did in their original position, enabling us to selectively induce SC on specific regions of the graphene surface. The manipulation procedure is precise, versatile, and reproducible. Lead islands can be displaced over distances of several hundredths of nanometres, in any direction, with nanometre precision. The manipulation process can be repeated several times, and the islands move across most of the inhomogeneities found in the graphene surface, such as step edges, grain boundaries, point defects etc. Importantly, several Pb islands can be consecutively manipulated to build complex SC-graphene hybrid structures. To illustrate this, we present in Fig 2d-f one example of such a multiple manipulation, where, as indicated by green arrows, five SC Pb islands have been consecutively approached to a selected graphene spot, outlined by a cross. This enables us, for example, to induce robust SC in this specific graphene spot, as evidenced in Fig. 2f, where a comparison of STS data measured before and after the collective manipulation shows the development of sharp coherence peaks and a lower in-gap local density of states.

The controlled manipulation of the position of lead islands opens a new playground to investigate S-N-S hybrid structures, where the specific configuration of the superconducting leads become a tuneable parameter. This enables the exploration of superconductivity induced in exactly the same normal region, here graphene, as a function of many different configurations of the superconducting leads. To our knowledge, there is no other experimental approach that allows for the successive engineering of superconducting junctions with such a degree of control. In the following, we show a series of experiments that illustrate the potential of this method.

In Figure 3a, we show a S-Graphene-S junction constructed by approaching the parallel edges of two Pb islands rotated 180º one respect to the other. The islands induce superconductivity in uncovered graphene and, in Figure 3b, we studied its spatial evolution through *dI/dV* spectra measured along a line perpendicular to both island edges. Using tip-induced manipulation, we approached one of the Pb islands to gradually reduce the distance between them from L=107 nm to L=44 nm, which allowed us to probe the evolution of the proximitized graphene region between the islands as a function of the spacing between them (Figure 3b). Interestingly, spectra measured close to each island differ, which implies that the islands induced superconductivity in a different way. For the largest spacing between islands studied, L=107 nm, two pairs of coherence peaks appear between the islands, the ones at larger bias decaying from the top island, while those with lower bias emerging stronger from the bottom island (left panel in Figure 3b). The evolution of the STS data along the region between the two islands can be reproduced by our simulations by considering that the Pb islands proximitized differently the graphene regions underneath. In Figure 3c, we show the simulation of an asymmetric $S_1$-Graphene-$S_2$ junction modelled as two superconducting regions with different gap, i.e. with $\Delta_{G1}$ = 1.2 meV and $\Delta_{G2}$ = 0.6 meV for the top and bottom Pb islands respectively, and using $D$ = 105 cm$^2$/s. For the closest spacing between islands studied, L=44 nm, the experiments indicate that the distinct proximity gaps unify into one single gap, distributed homogeneously across the graphene region (right panel in Figure 3b).

These studies can be extended two-dimensional arrangements of SC structures, thus increasing the complexity of the proximitized Pb-graphene systems. In Figure 3d, we show a S-N-S junction formed by two parallel Pb islands, which proximitize the region between them, as revealed by the green green *dI/dV* plot in Figure 3f. Subsequently, we moved a third Pb island to shunt the right-hand side of the S-N-S junction, forming a semi-open 2D corral (Figure 3e). STS data in the same spot as before show an enhancement of the induced superconductivity (red curve in Figure 3f), even though this region lies far nm away from the shunted region. By performing simulations on two-dimensional SC island arrangements, our experiments can be rationalized as due to a change in the boundary conditions (Figures. 3g-i).

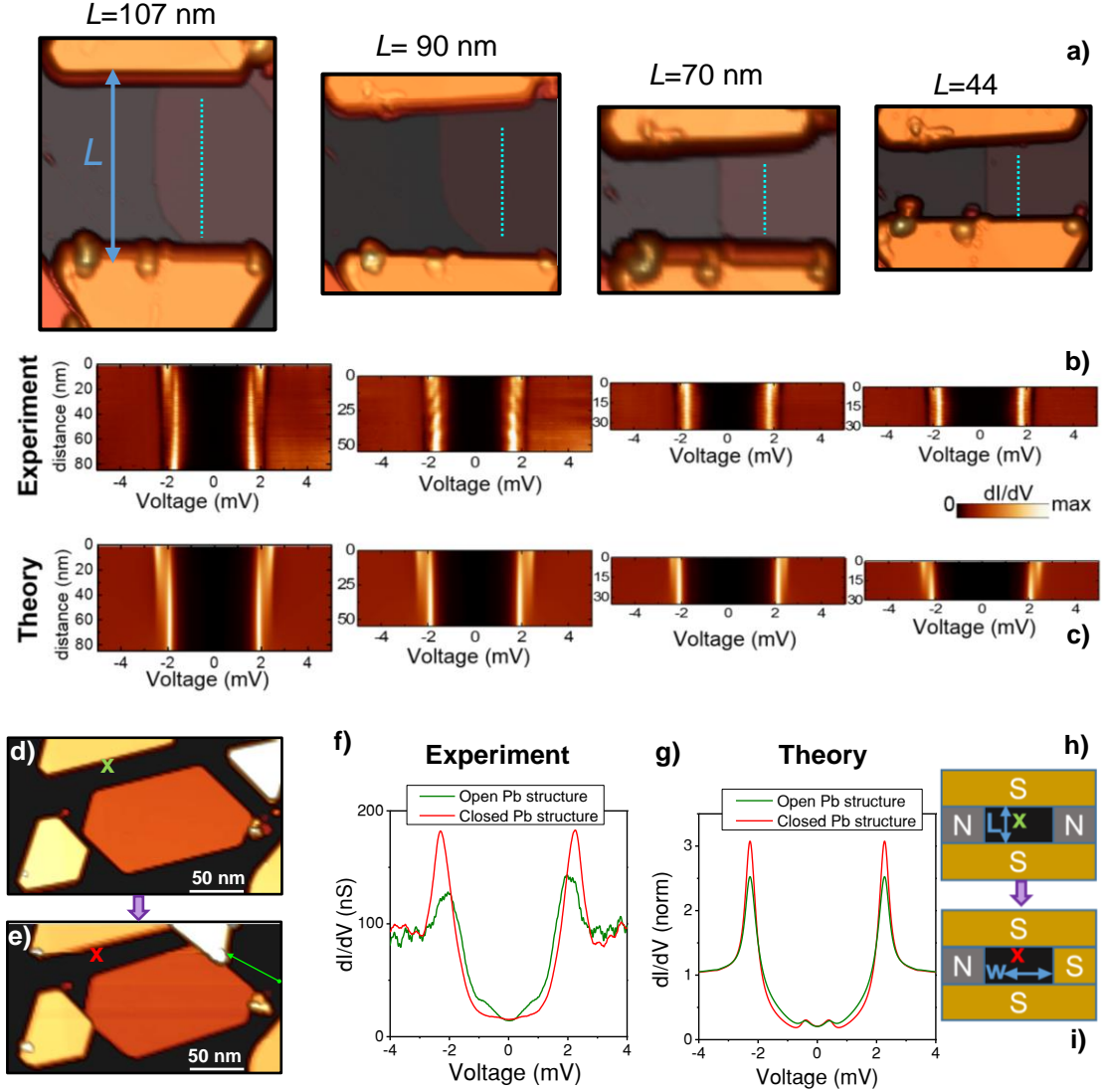

**Figure 3. Pb island manipulation as a tool to investigate SNS junctions.** a) STM images showing the same S-Graphene-S junction, formed by two Pb islands with parallel edges, whose separation $L$ is tuned by displacing the Pb islands with the STM tip. b) Corresponding experimental conductance maps [$dI/dV(x,E)$] along the blue dotted lines in (a); $V$bias = 10 mV; $I$set = 0.5 nA. $T$=1.3K. c) Calculated conductance maps [$dI/dV(x,E)$], considering two superconductors with different gaps $\Delta_{G1}$ = 1.2meV and $\Delta_{G2}$ = 0.6 meV to simulate the proximitized graphene regions below the top and bottom Pb islands respectively d-e) STM images showing the transition from a 1D S-Graphene-S junction to a semi-open 2D junction. f) dI/dV curves measured on the same graphene spot outlined by crosses in d) and e) before (green curve) and after (red curve) the manipulation. g-i) 2D simulations to reproduce the experimental results from d-f). We have considered rectangular domains where the diffusive normal region has a length $L$ =70 nm and a width $W$=100 nm. In these domains the normal graphene region is coupled to reservoirs that can be either normal or superconducting, depending on the type of corral. $\Delta_G$ = 0.95 meV; $\Gamma_G$ = 0.15 meV; $\xi$ =70 nm.

A final step forward in the manipulation of 2D SN hybrid structures is the construction of closed loops, formed by interconnected Pb islands completely enclosing graphene regions, i.e., graphene superconducting nanocorrals. In Figure 4a we show an example of such a superconducting corral, formed after the controlled manipulation of several Pb islands. STS data, measured along the blue dotted line outlined in Figure 4a, shows a strong and constant proximity-induced superconductivity in the whole enclosed graphene region (Figure 4b). Our 2D simulation nicely matches the homogeneous proximitized region inside the corral (Figure 4c). These superconducting graphene corrals are very interesting because the superconductivity induced in graphene can be comparable to the one of bulk Pb (Supporting Figure S4). Additionally, the SC LDOS is essentially constant within the graphene SC corral, as long as the dimensions of the corral are comparable to the graphene $\xi$, which, as shown before, for graphene on SiC(000-1) amounts to $\xi \approx 70$ nm (Figure 4c). Importantly, the SC corrals as a whole can be moved using the same tip manipulation procedure than for the single islands, which can be exploited to induce a well-defined superconductivity in a specific graphene region or even to use the SC corral as a local probe for graphene magnetism, as we demonstrate next.

While pristine graphene is not magnetic, local magnetic moments can be generated on it by atomic vacancies, H atoms, or grain boundaries. In every case, magnetism is caused by an unbalance in the number of $\pi$-electrons on each graphene sublattice, in line with Lieb's theorem [39]. Although there is a vast number of theoretical works on this subject, the experimental detection of graphene magnetism is challenging [40-43]. Here, demonstrate that superconducting corrals can be used as spin detectors in graphene, by placing them around a defect and detecting the effect of unpaired spins on the proximity-induced SC-gap of graphene. Localized magnetic moments interacting with correlated pairs in the proximitized region induce narrow excitations, known as Yu-Shiba-Rusinov states (YSR), inside the proximitized gap [44-48]. Thus, the presence of YSR states, which can be detected by STM, is an unequivocal proof of the existence of local magnetic moments, as shown in a large variety of systems [17, 32, 47-55]

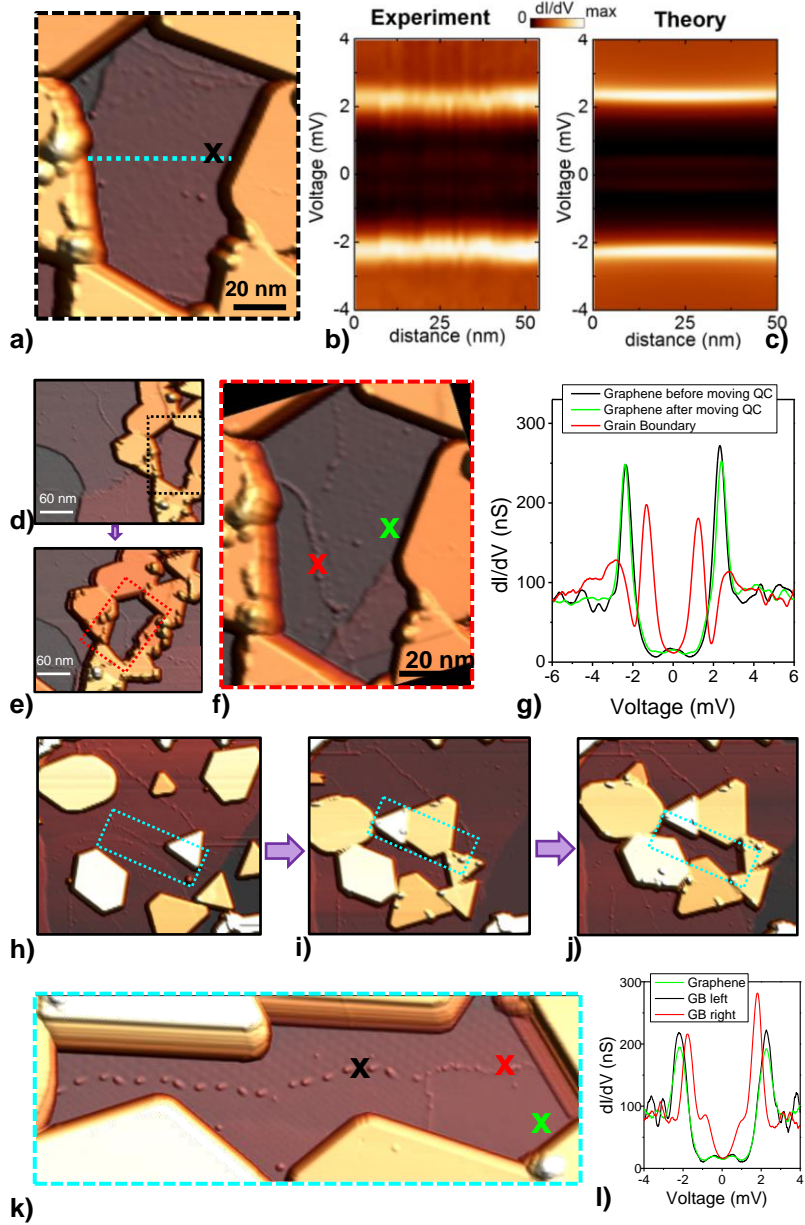

**Figure 4. Exploiting SC graphene nanocorrals as local experimental probes of graphene magnetism.** a) STM image showing a graphene SC nanocorral, i.e. a graphene region completely enclosed by Pb islands, built on SiC(000-1). b) Experimental conductance map [$dI/dV(x,E)$] along the blue dotted line in (a); $V_{bias}$ = 10 mV; $I_{set}$ = 0.5 nA. c) Calculated conductance map [$dI/dV(x,E)$], with $\Delta_G$ = 1.20 meV; $\Gamma_G$ = 0.20 meV; $\Gamma_{sf}$=0.2; $\xi$=70 nm. d) Large scale STM image with the nanocorral of (a) outlined by a black dotted rectangle. e) STM image showing the same region after the manipulation of the whole SC nanocorral, dotted red rectangle, to accommodate a graphene grain boundary. f) STM image of the SC nanocorral after the manipulation, where the presence of the grain boundary is clearly appreciated. g) $dI/dV$ curves measured inside the SC nanocorral. In black, STS data measured in the pristine graphene region outlined by a black cross in (a). In green, STS data measured in an equivalent pristine graphene position inside the SC corral after the manipulation, green cross in (f). In red, STS data measured on the graphene grain boundary, red cross in (f), showing the emergence of in-gap YSR states, which demonstrates that this pure carbon defect induces local magnetic moments in graphene. h-k) Formation and lateral manipulation of a SC nanocorral to accommodate a grain boundary between 3 graphene domains. l) *dI/dV* curves measured inside the SC nanocorral in (k). While red curve, measured on the right-side grain boundary, shows YSR states due to the existence of local magnetic moments on it, black curve is barely distinguishable from the one measured on pristine graphene, indicating the absence of magnetic moments on this specific grain boundary.

As we mentioned before, the superconducting Pb corrals can be moved as a whole, inducing superconductivity in the selected regions with reproducible properties. Therefore, to demonstrate the validity of the corrals as a sensor for local magnetism, we moved them to enclose specific graphene regions with potential to host local magnetic moments, and searched for the existence or absence of YSR in-gap states in the proximitized region.

As pictured in Figs. 4d-f, we moved the superconducting corral of Figure 4a to a distance 100 nm away, to now enclose a graphene region with a grain boundary (compare resulting image in Fig. 4f with the initial location in Fig. 4a ). STS plots on defect-less graphene regions inside the SC corral, far away from any defect, the superconductivity is well induced and the LDOS is the same as it was before the SC corral manipulation (black and green *dI/dV* curves in Figure 4g). On the other hand, *dI/dV* spectra measured on top of the grain boundary (red curve in Figure 4g) reveals the emergence of YSR in-gap states, demonstrating the presence of graphene magnetic moments induced by this grain boundary[17]. Due to the reproducibility and high degree of control of the manipulation process, this technique is versatile. In Figure 4h-j, we show the complete process for the formation and controlled displacement of a SC corral built, by manipulating seven SC Pb islands, with an elongated shape to be able to enclose three different graphene domains, separated by the corresponding grain boundaries (Figure 4k). As shown in the *dI/dV* plots of Figure 4l, this method allows to identify which grain boundaries host magnetic moments (red plot) and which ones do not (black plot).

The manipulation of the Pb islands with the STM tip worked with 100% reproducibility in all the graphene systems that we have investigated, which enabled us, for example, to create SC nanocorrals in all of them. In Supporting Figure S5 we show the results on HOPG where the coherence length of the induced SC is only $\xi \approx 10nm$. Such shorter coherence length is clearly reflected in the spatial evolution of the induced SC inside the corrals, which is not further constant for a nanocorral of only 25nm width (Supporting Figure S5). The high reproducibility of our method for a large variety of decoupled graphene systems, suggest that it should also work on exfoliated graphene sheets, where the graphene density of charge carriers can be selectively tuned by external electrostatic gating, providing and additional channel to tune graphene SC. This also gives the possibility to study the interplay between induced superconductivity and other proximity effects effects (SOC [37], magnetism [38]) generated in graphene by an underlying layer.

## 3. Conclusion

To summarize, our work provides a new avenue for exploring and tuning graphene superconductivity. With the toolbox we have developed, the playground is now ready for the incorporation of many additional ingredients, such as magnetic fields, Josephson currents, strain, twisted bilayers, SC islands of different materials, external electronic gating, etc. This should enable to visualize Josephson vortices, analyse the interplay of pseudomagnetic fields and SC, detect graphene magnetism (e.g., in twisted bilayers), generate new electronic phases, and to reach a much more comprehensive understanding of graphene superconductivity and of S-N-S junctions in general.

## 4. Experimental Section/Methods

*Sample preparation and experimental details.*

To ensure an atomistic control of the samples, we performed all preparation procedures and measurements under UHV conditions. During the whole process -imaging the pristine graphene sample => deposition of Pb on it => back imaging and manipulation of the Pb island- the sample was kept in the same UHV system.

We grew the multilayer graphene on a 6H-SiC(000-1) sample (C face), following the method described in [56]. The process consists in heating, at 1600°C in a RF furnace, the SiC substrate held in a graphite crucible under an Ar atmosphere (1 bar Ar) for 30 min. Prior to this graphitization step, the substrate is etched in an Ar/$H_2$ mixture [57]. After the growth, the sample was transferred into a separate ultrahigh vacuum (UHV) setup and outgassed. The surface graphene layer features single-crystalline domains 100-500 nm wide with different crystallographic orientations [58]. In this system, the rotational disorder of the graphene layers electronically decouples π bands and the pristine graphene surface layer is essentially neutral, presenting a very low electron doping [29, 59]. We grew ML and BL graphene on a 6H-SiC(0001) substrate by thermal desorption of silicon at high temperatures, in ultra-high vacuum (UHV) [60]. In this system, graphene layers are electron-doped, with Dirac energy ($E_D$) 0.45 eV and 0.30 eV below the Fermi energy ($E_F$) for ML and BL respectively. Highly ordered pyrolytic graphite (HOPG) surfaces, which present the AB Bernal stacking, were obtained by the in-situ exfoliation and further annealing at 650°C of HOPG samples [40].

In all these substrates, to grow SC Pb islands, we deposited 1-3ML of Pb at a rate of 0.5-2 ML/min while keeping the sample at room temperature. This formed triangular and hexagonal Pb islands randomly distributed on the graphene surface, with heights between 2-10 nm and sides between 20-300 nm, see Figure 1 and Supporting Figures. S1-2 [30,17].

*dI/dV* conductance curves were obtained by the numerical differentiation of *I-V* curves. We used a home-made UHV-LT-STM with a base sample and tip temperature that are $T_{sample} = 4.0$ K, and $T_{tip} = 3.1$ K respectively, and a SPECS GmbH STM, operative at 1.2 K under UHV conditions, all temperatures being way below the Pb superconducting Tc = 7.2 K. We use Pb SC tips to improve the resolution beyond the thermal limit [31, 32]. The SC of the tips is essentially the Pb bulk one, as we directly calibrate in large Pb islands [17].

All the STM data were measured/processed using the WSxM software [61].

*Theoretical description of the proximity effect in graphene: Usadel equations*

Our theoretical description of the proximity superconductivity in graphene is based on the so-called Usadel equations that summarize the quasi-classical theory of superconductivity in the diffusive limit [62]. In this limit the mean free path is assumed to be much smaller than the superconducting coherence length. Inside the graphene region in between the island, this coherence length is given by $\xi = \sqrt{\hbar D/\Delta_{Gi}}$, where $D$ is graphene's diffusion constant and $\Delta_{Gi}$ is the energy gap in the graphene region below the superconducting leads. Within the quasi-classical theory, all the equilibrium properties are described in terms of a momentum averaged retarded Green's function $\hat{G}(\vec{R}, E)$, which depends on position $\vec{R}$ and energy $E$. This propagator is indeed a 2 × 2 matrix in electron-hole space (indicated by a caret)

$$\hat{G} = \begin{pmatrix} g & f \\ \tilde{f} & \tilde{g} \end{pmatrix}. \tag{1}$$

In this work we are particularly interested in the local density of states, which can easily computed form the knowledge of the Green's function as

$$\rho(\vec{R}, E) = -\frac{1}{\pi} \text{Im}\{g(\vec{R}, E)\} . \tag{2}$$

Our goal here is to describe the proximity effect in the graphene regions between and close to the Pb islands, which effectively form different kinds of superconductor-normal (*SN*) junctions. Ignoring inelastic interactions, the propagator $\hat{G}(\vec{R}, E)$, satisfies the stationary Usadel equation, which in the normal regions of our junctions reads[1]

$$\frac{\hbar D}{\pi} \nabla(\hat{G}\nabla\hat{G}) - \frac{\Gamma_{sf}}{2\pi}[\hat{\tau}_3 \hat{G} \hat{\tau}_3, \hat{G}] + E[\hat{\tau}_3, \hat{G}] = 0 , \tag{3}$$

where $\hat{\tau}_3$ is the Pauli matrix in electron-hole space and $\Gamma_{sf}$ is a spin-flip scattering rate (with dimensions of energy) that describes pair-breaking mechanism such as magnetic impurities. Equation (3) must be supplemented by the normalization condition $\hat{G}^2 = -\pi^2$. To solve numerically the Usadel equation, it is convenient to use the so-called Riccati parameterization[63], which accounts automatically for the normalization condition. In this case, the retarded Green's functions are parameterized in terms of two coherent functions $\gamma(\vec{R}, E)$ and $\tilde{\gamma}(\vec{R}, E)$ as follows

$$\hat{G} = -\frac{i\pi}{1+\gamma\tilde{\gamma}} \begin{pmatrix} 1 - \gamma\tilde{\gamma} & 2\gamma \\ 2\tilde{\gamma} & \gamma\tilde{\gamma} - 1 \end{pmatrix}. \tag{4}$$

Using their definition in Eq. (4) and the Usadel equation (3), one can obtain the transport equations for these functions in the normal region [64,65]. In the case of the description of the proximity effect close to a single Pb island or in the region between two islands, we have modelled the system as a one-dimensional (1D) *SN* or *SNS* junction, respectively. In this case the equations for the coherent functions read

$$\partial_x^2 \gamma + \frac{\tilde{f}}{i\pi}(\partial_x \gamma)^2 - 2\left(\frac{\Gamma_{sf}}{E_T}\right)\gamma \frac{\tilde{g}}{i\pi} + 2i\left(\frac{E}{E_T}\right)\gamma = 0 , \quad (5)$$

$$\partial_x^2 \tilde{\gamma} + \frac{f}{i\pi}(\partial_x \tilde{\gamma})^2 + 2\left(\frac{\Gamma_{sf}}{E_T}\right)\tilde{\gamma} \frac{g}{i\pi} + 2i\left(\frac{E}{E_T}\right)\tilde{\gamma} = 0 . \quad (6)$$

Here, $x$ is the dimensionless coordinate that describes the position along the $N$ region and it ranges from 0 (left lead) to 1 (right lead) and $E_T = \hbar D/L^2$ is the Thouless energy of a normal region of length $L$. The expressions for $g, f, \tilde{g}, \tilde{f}$ are obtained by comparing Eq. (1) with Eq. (4). Notice that, in general, Eqs. (5) and (6) couple the functions with and without tilde. Now, we must provide the boundary conditions for these two equations. For interfaces with perfect transparency, as we assumed here, such conditions at the ends of the $N$ region result from the continuity of the Green's functions at the interfaces. Thus, the coherent functions in the interface with a superconducting region of gap $\Delta$ adopts the form:

$$\gamma = \gamma_S(E) = -\frac{\Delta e^{i\varphi_S}}{E^R + i\sqrt{\Delta^2 - (E^R)^2}} \quad \text{and} \quad \tilde{\gamma} = \tilde{\gamma}_S(E) = \frac{\Delta e^{-i\varphi_S}}{E^R + i\sqrt{\Delta^2 - (E^R)^2}} , \quad (7)$$

where $E^R = E + i0^+$ and $\varphi_S$ is the superconducting phase, which we shall assume to be zero in all the calculations presented here. In the case of a normal reservoir/lead the coherent functions are set to zero.

For the description of the proximity effect in the corrals (both open and closed), we have considered rectangular domains where the diffusive normal region has a length $L$ and a width $W$. In these domains the normal graphene region is coupled to reservoirs that can be either normal or superconducting, depending on the type of corral. The normal domain (normal graphene region) is assumed to lie in the *xy*-plane, where $x \in [0, L]$ and $y \in [-W/2, W/2]$. If we introduce the dimensionless coordinates $\tilde{x} \in [0,1]$ and $\tilde{y} \in [-1/2, 1/2]$, the coherent function $\gamma(\vec{R}, E)$ satisfies the following equation (in the absence of spin-flip scattering)

$$\partial_{\tilde{x}}^2 \gamma + \left(\frac{L}{W}\right)^2 \partial_{\tilde{y}}^2 \gamma + \frac{\tilde{f}}{i\pi}\left[(\partial_{\tilde{x}}\gamma)^2 + \left(\frac{L}{W}\right)^2 (\partial_{\tilde{y}}\gamma)^2\right] + 2i\left(\frac{E}{E_T}\right)\gamma = 0 . \quad (8)$$

There is another equation (coupled to this one) for $\tilde{g}$ that can be obtained from Eq. (8) by exchanging $g$ by $\tilde{g}$ (and vice versa). Finally, we need to specify the boundary conditions for these coherent functions. Again, we have assumed perfect transparent interfaces and thus, the coherent functions adopt their bulk values in those interfaces. That means that they are zero if the reservoir is normal (open end of a corral) and they adopt the form in Eq. (7) is the corresponding reservoir is a superconductor (in our case a Pb island).

In all cases the set of coupled second-order nonlinear differential equations satisfied by the coherent functions, see Eqs. (5,6,8), were solved numerically using the so-called relaxation method for boundary value problems[66], which was adapted in our case to deal with partial differential equations. The technical details have been described elsewhere [65,67]. With the solution of these equations, we computed the corresponding local density of state, see Eq. (2), which was finally translated into a differential conductance using the standard tunnelling formula to directly compare with our experimental results.

**Supporting Information**
Supporting Information is available from the Wiley Online Library or from the author.

**Acknowledgements**
We acknowledge funding from the Spanish Ministry of Science and Innovation MCIN/AEI/10.13039/297 501100011033 though grants # PID2020-115171GB-I00, PID2020-114880GB-I00, PID2019-107338RB-C61 and the "María de Maeztu" Programme for Units of Excellence in R&D (CEX2018-000805-M, CEX2020-001038-M), the Comunidad de Madrid NMAT2D-CM program under grant S2018/NMT-4511, the Comunidad de Madrid, the Spanish State and the European Union by the Recovery, Transformation and Resilience Plan "Materiales Disruptivos Bidimensionales (2D)" (MAD2D-CM)-UAM3 and the European Union through the Next Generation EU funds and the Horizon 2020 FET-Open project SPRING (No. 863098). J. C. C. thanks the German Science Foundation DFG and SFB 1432 for sponsoring his stay at the University of Konstanz as a Mercator Fellow.

**TOC**

We present a novel method, based on the proximity effect in combination with STM manipulation, which enables to introduce superconductivity at will in any graphene region. This represents a crucial breakthrough, both to shed light in the fundamental understanding of superconductivity in graphene-based systems, and also to build functional superconducting-graphene hybrid structures.

*Eva Cortés-del Río, Stefano. Trivini, José I. Pascual, Vladimir Cherkez, Pierre Mallet, Jean-Yves Veuillen, Juan. C. Cuevas, Iván Brihuega\**

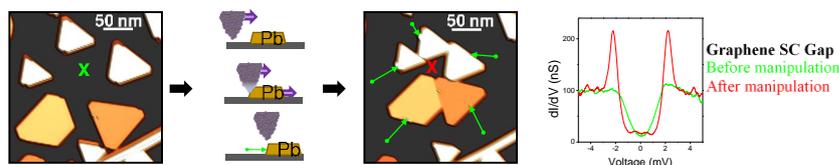

Title **Shaping graphene superconductivity with nanometer precision**

**ToC figure**

# Supporting Information

**Title: Shaping graphene superconductivity with nanometer precision**

*Author(s), and Corresponding Author(s)\**

*Eva Cortés-del Río, Stefano. Trivini, José I. Pascual, Vladimir Cherkez, Pierre Mallet, Jean-Yves Veuillen, Juan. C. Cuevas, Iván Brihuega\**

**Supporting Data**

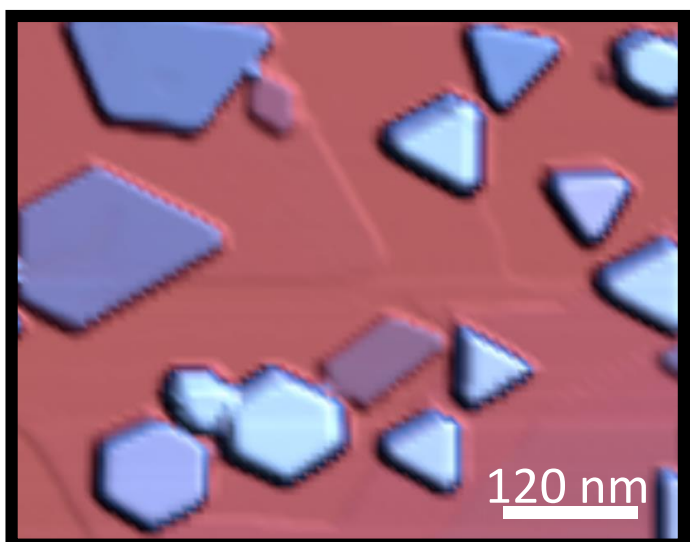

**Figure S1. Pb islands on graphene on ML and BL graphene on SiC(0001).**
Large scale STM image showing the general morphology of the samples after RT Pb deposition Temperature = 4.2K; Vb = 600 mV; It = 0. 5 nA.

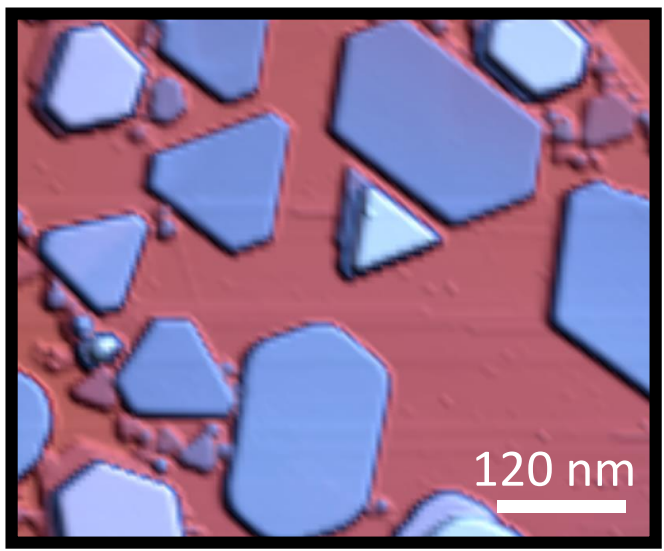

**Figure S2**. **Pb islands on HOPG.**
Large scale STM image showing the general morphology of the samples after RT Pb deposition. Temperature = 4.2K; Vb = 90 mV; It = 0. 5 nA.

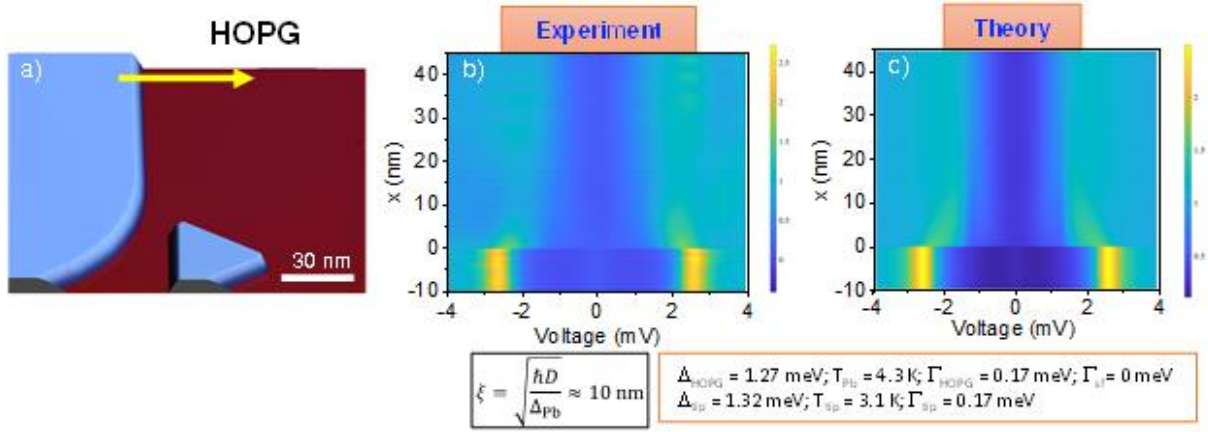

**Figure S3. Inducing SC in HOPG by proximity.** . a) STM image showing a 6nm height Pb island on top of HOPG. b) Experimental conductance map [d*I*/d*V*(*x*,*E*)] along the yellow line in (a), ), showing the evolution of SC as a function of *x*, the distance from a the Pb island. Negative *x* means *dI/dV* spectra measured on top of the Pb island (*V*bias = 10 mV; *I*set = 0.5 nA). c) Calculated conductance map [d*I*/d*V*(*x*,*E*)].

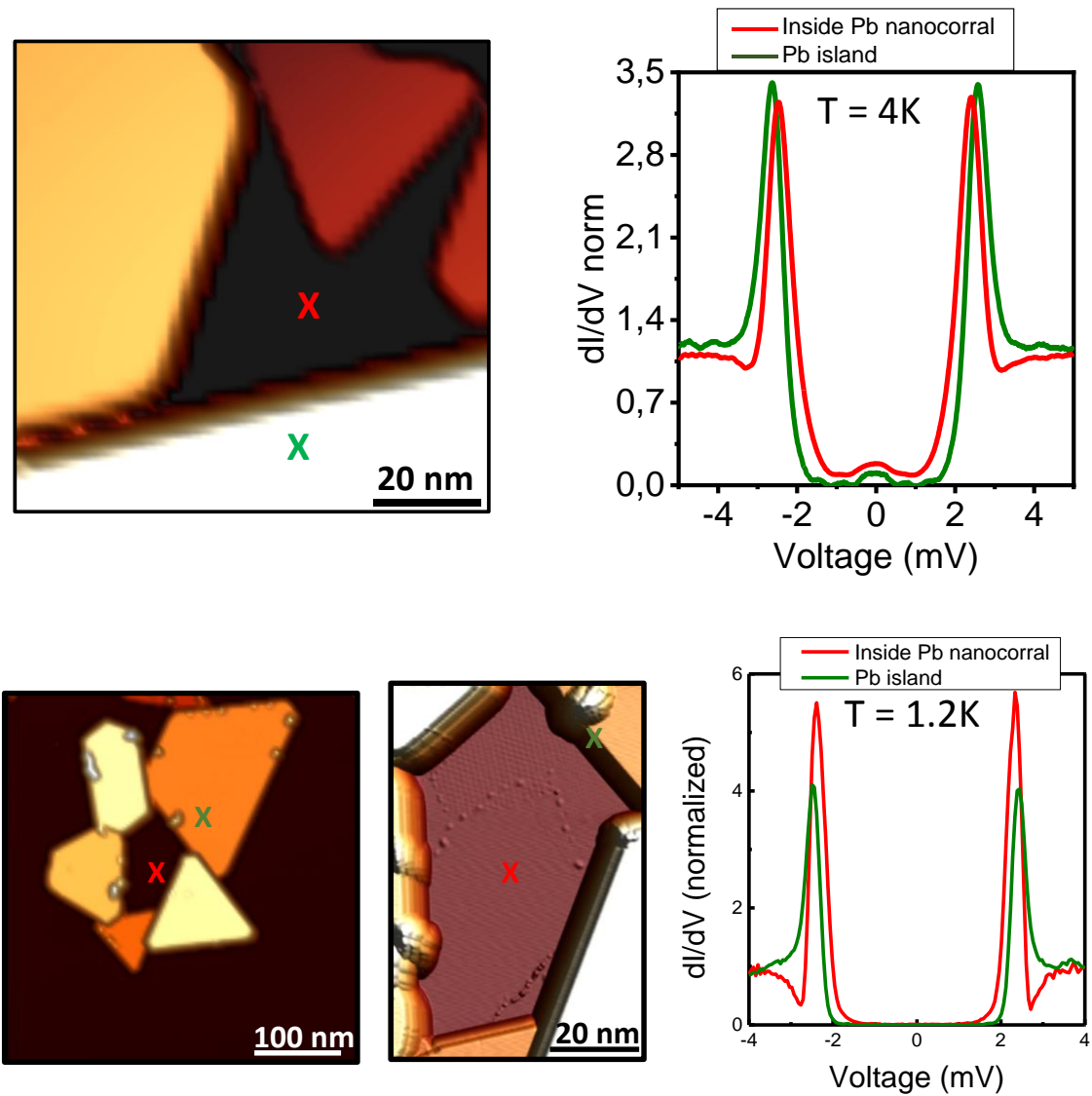

**Figure S4.** Superconductivity inside a SC graphene nanocorral. Our STS data, shows that the superconductivity induced inside the Pb nanocorrals is similar to the one of bulk Pb, presenting a hard gap. Top and bottom pannels show the comparison between dI/dV spectra measured on the Pb islands and inside Pb nanocorrals, on the position marked by crosses, at 4K and 1.2K respectively.

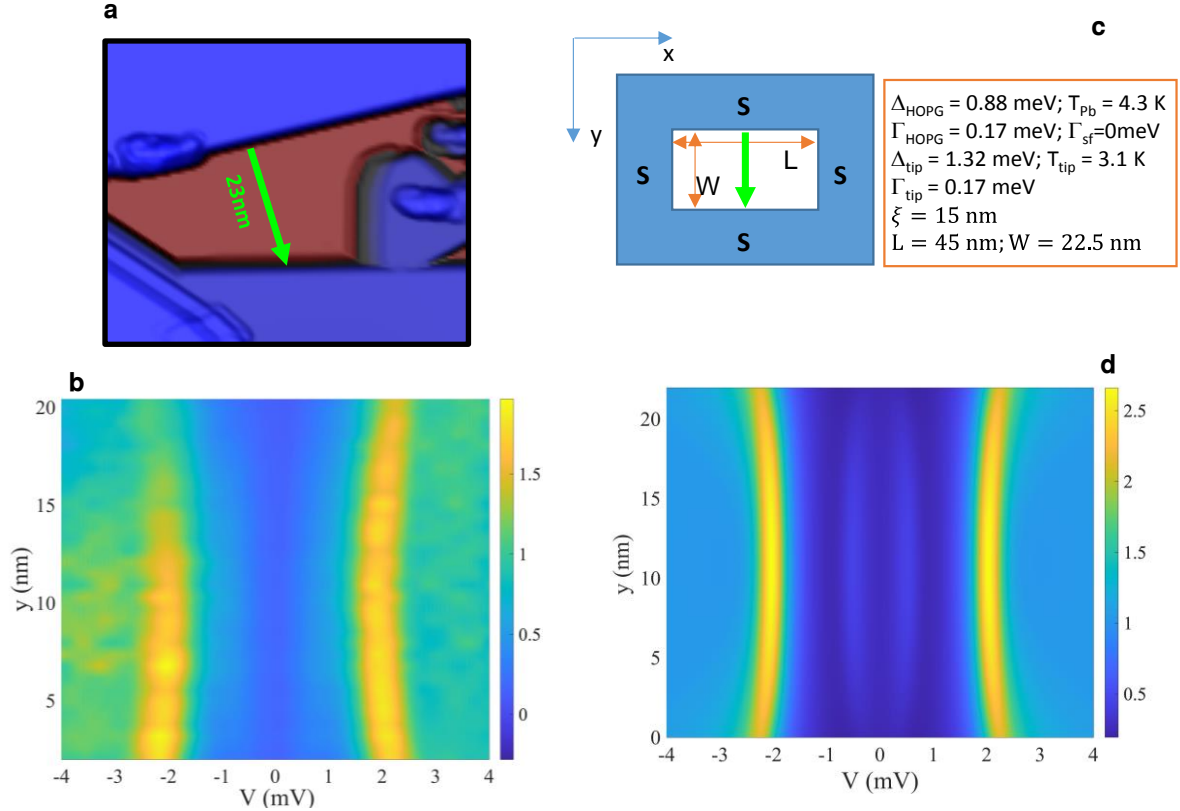

**Figure S5. SC graphene nanocorral on HOPG**

a) STM topograph of a closed SC corral on HOPG. $V_b$ = 90 mV, $I_t$ = 0.05 nA. **b)** STS data along the blue arrow in a). STS spectra along the corral reveals a slight decay in the SC gap as we move into graphene, which is recovered as we approach the Pb island again. **c)** Simulated rectangular corral of graphene surrounded by a superconductor. **d)** Calculated spectra for a corral with the similar dimensions to the one in a). The calculations reproduce well our experiment with a coherence length of $\xi$ = 15 nm, consistent with the result previously shown for the isolated island (Extended Data Fig. 4).